\documentclass[prd,twocolumn,floatfix,superscriptaddress]{revtex4}
\usepackage{graphicx}
\usepackage{epsfig}
\usepackage{bm}
\usepackage{amssymb}
\usepackage{float}
\usepackage{amsmath}
\usepackage{dcolumn}
\usepackage{cancel}
\usepackage[colorlinks]{hyperref}
\usepackage[usenames,dvipsnames]{color}
\hypersetup{
     breaklinks=true,
    pdfstartview={FitH},    
    colorlinks=true,       
    linkcolor=blue,          
    citecolor=red,        
    filecolor=magenta,      
    urlcolor=blue,           
    anchorcolor=green,      
    linktocpage=true
}

\def\doi{http://doi.org}
\def\arxiv{http://arxiv.org/abs}
 
 \def\e{\mathrm{e}}

\begin{document}

\title{Holographic inflation}

 \author{Shin'ichi Nojiri}
\affiliation{Department of Physics, Nagoya University, Nagoya 464-8602, Japan}
\affiliation{Kobayashi-Maskawa Institute for the Origin of Particles and the
Universe, Nagoya University, Nagoya 464-8602, Japan}

\author{Sergei D. Odintsov}
\affiliation{Institut de Ciencies de lEspai (IEEC-CSIC), 
Campus UAB, Carrer de Can Magrans, s/n 
08193 Cerdanyola del Valles, Barcelona, Spain}
\affiliation{Instituci\'{o} Catalana de Recerca i Estudis Avan\c{c}ats
(ICREA), Passeig Llu\'{i}s Companys, 23 08010 Barcelona, Spain}

\author{Emmanuel N. Saridakis}
\affiliation{Department of Physics, National Technical University of Athens, Zografou
Campus GR 157 73, Athens, Greece}
\affiliation{Department of Astronomy, School of Physical Sciences, University of Science 
and Technology of China, Hefei 230026, P.R. China}

\begin{abstract}
We apply the holographic principle at the early universe, obtaining an 
inflation realization of holographic origin. Such a consideration has equal footing with 
its well-studied late-time application, and moreover the decrease of the horizons at 
early times naturally increases holographic energy density at inflationary scales. 
Taking as Infrared cutoff the particle or future event horizons, and adding a simple 
correction due to the Ultraviolet cutoff, whose role is non-negligible at the high energy 
scales of inflation, we result in a holographic inflation scenario that is very 
efficient in incorporating inflationary requirements and predictions. We first
 extract analytically  the solution of the Hubble function  in an implicit 
form, which   gives a scale factor evolution of the desired e-foldings.
Furthermore, we analytically calculate the Hubble slow-roll parameters and then the 
inflation-related observables, such as the scalar spectral index and its running, 
the tensor-to-scalar ratio, and the tensor spectral index.
 Confronting the predictions with   Planck 2018 observations we 
show that the agreement is perfect and in particular 
deep inside the 1$\sigma$ region.  
\end{abstract}

\maketitle

\emph{Introduction} -- 
The holographic principle originates from black hole thermodynamics and string theory 
\cite{tHooft:1993dmi,Susskind:1994vu,Witten:1998qj,Bousso:2002ju}, and establishes a 
connection  of the Ultraviolet cutoff of a quantum field theory, which is related to 
the vacuum energy, with the largest distance of this theory, which is related to 
causality and the  quantum field theory  applicability at large distances 
\cite{Cohen:1998zx}. This consideration has been applied extensively at a cosmological 
framework at late times, in which case the  obtained 
vacuum energy constitutes  a dark energy sector of holographic origin, called 
holographic dark energy \cite{Li:2004rb} (for a review see \cite{Wang:2016och}).
In particular, the holographic energy density is proportional to 
the inverse squared Infrared cutoff $ L_\mathrm{IR}$, which since     is related to 
causality it must be a form of horizon,
namely
 \begin{equation}
 \label{basic}
\rho=\frac{3c^2}{\kappa^2 L^2_\mathrm{IR}},
\end{equation}
with $\kappa^2$ the gravitational constant and $c$ a parameter. 
Holographic dark energy proves to have interesting phenomenology, both in 
its basic 
\cite{Li:2004rb,Wang:2016och,Enqvist:2004xv,Zhang:2005yz,
Elizalde:2005ju,Pavon:2005yx, Guberina:2005fb,  Nojiri:2005pu,
Kim:2005at} as well 
as in its various extensions 
\cite{Ito:2004qi,Gong:2004cb,Saridakis:2007cy,
Gong:2009dc,BouhmadiLopez:2011xi,Malekjani:2012bw,Khurshudyan:2014axa,
Landim:2015hqa}, and it can fit   
observations
\cite{Huang:2004wt,Zhang:2005hs,Li:2009bn,Feng:2007wn,Zhang:2009un,Lu:2009iv,
Micheletti:2009jy}.

Despite the extended research on the application of holographic principle in late-time 
cosmology and dark-energy epoch, there has not been any attempt in applying it at early 
universe, namely to obtain an inflationary realization of holographic origin. 
Nevertheless, such consideration has equal footing with its late-time application, and 
moreover, observing the form of (\ref{basic}), we deduce that since at early times the 
largest distance is small, the holographic energy density is naturally suitably 
large in order to lie in the inflationary scale.

In the present Letter we are interested in investigating holographic inflation, namely to 
acquire a successful inflation triggered by the energy density of holographic origin. 
Since the involved energy scales at this epoch are high, we should additionally consider 
a correction coming from the Ultraviolet cutoff. As we show, although the basic scenario 
is very simple and natural, it can be very efficient and results to inflationary 
observables in perfect agreement with observations. Furthermore, as we show, one can 
extend the basic scenario to more subtle constructions.\\


\emph{Holographic inflation} -- 
In this section we will  construct the basic model of holographic inflation.
We consider a homogeneous and isotropic Friedmann-Robertson-Walker (FRW)
 geometry with metric
\begin{equation}
\label{FRWmetric}
ds^2=- dt^2+a^2(t) \left(\frac{dr^2}{1-kr^2}+r^2 d\Omega^2\right),
\end{equation}
where $a(t)$ is the scale factor and with
$k=0,+1,-1$ corresponding respectively to flat, close and open spatial geometry. In this 
work  for convenience we focus on the flat case, nevertheless the 
generalization to non-flat geometry is straightforward.   

In a general inflation scenario the first Friedmann equations is written as
\begin{equation}\label{FR1}
H^2=\frac{\kappa^2}{3} \rho_\mathrm{inf},
\end{equation}
where $\rho_\mathrm{inf}$ is the  energy density of the (effective) fluid that drives 
inflation, which can originate from a scalar field, from modified-gravity, or from 
other sources and mechanisms. Note that as usual we have neglected the contributions from 
other components, such as the matter and radiations sectors.

In this work we consider that inflation has a holographic origin, namely that 
its source is the holographic energy density. Hence, imposing that  
$\rho_\mathrm{inf}$ is $\rho$ of (\ref{basic}), the Friedmann equation (\ref{FR1}) 
for an expanding universe becomes simply  
\begin{equation}
\label{H2}
H=\frac{c}{L_\mathrm{IR}}\, .
\end{equation} 
As it is well-known, ${H}^{-1}$ corresponds to the radius of the cosmological 
horizon which gives the limit of the causal relations from the viewpoint of the classical 
geometry, namely any information beyond the horizon cannot affect us at least at the 
time when we measure $H$. The original idea of the holographic energy paradigm comes by 
identifying 
the classical horizon radius with the infrared cutoff in the quantum field theory, 
since we neglect any contribution coming from energy scales smaller than the cutoff 
scale. This implies that we may relate, as we will see later, the horizon radius 
 ${H}^{-1}$ with the Infrared cutoff, since the information in energy scales larger 
than the Ultraviolet cutoff scale is irrelevant.

 The simplest choice is that $L_\mathrm{IR}$ should be exactly the Hubble radius, which 
however 
cannot 
be used at late-times application since it cannot lead to 
an accelerating universe \cite{Hsu:2004ri}, and this is the case for   the 
next guess, namely the  particle horizon. Hence, one can use  
the future event horizon \cite{Li:2004rb}, the age of the universe or the conformal time
 \cite{Cai:2007us,Wei:2007ty}, the 
inverse square root of the Ricci curvature
\cite{Gao:2007ep},  a combination of Ricci, Gauss-Bonnet 
\cite{Saridakis:2017rdo} or other curvature invariants, or even consider a   
general  Infrared cutoff as an arbitrary function of the above and their derivatives
 \cite{Nojiri:2017opc}.

Contrary to the case of late-time application of holographic principle, namely the 
holographic dark energy,  in the present inflationary 
application almost all the above choices can be successful in driving inflation, due to 
the absence of matter sector, apart from the simplest case of the 
Hubble radius which leads to a trivial result.  In particular,
we may consider the particle horizon $L_\mathrm{p}$ 
or the future event horizon $L_\mathrm{f}$, which are given as
\begin{equation}
\label{H3}
L_\mathrm{p}\equiv a
\int_0^t\frac{dt}{a}\ ,\quad L_\mathrm{f}\equiv a \int_t^\infty \frac{dt}{a}\, .
\end{equation}
Inserting these into (\ref{H2}) we obtain  
\begin{equation}
\label{H5}
\frac{d}{dt}\left(\frac{c}{aH}\right)= \frac{m}{a}\, ,
\end{equation}
where $m=1$  corresponds to the particle horizon and $m=-1$ 
 to the future event horizon. 
In the second case, and for $c=1$, we immediately extract the de Sitter solution
\begin{equation}
\label{dS1}
a=a_0\e^{H_0 t}\, ,
\end{equation}
with $a_0$,$H_0$ the two integration constants. Hence, as we observe, the basic 
inflationary feature can be straightforwardly obtained.

Since we investigate the application of holographic principle at early times, and thus at 
high energy scales, apart from the Infrared cutoff we should consider the effects of the 
Ultraviolet cutoff $\Lambda_\mathrm{UV}$ too. In particular, at this regime the quantum 
effects become important, and thus the Infrared cutoff acquires a correction by the 
Ultraviolet one, which as was shown in 
\cite{Nojiri:2004pf} takes the simple form
\begin{equation}
\label{H8c}
L \equiv \sqrt{  L^2_\mathrm{IR} + \frac{1}{\Lambda_\mathrm{UV}^2}}\, .
\end{equation}
Therefore, inserting this corrected expression into (\ref{H2}), with $L_\mathrm{IR}$
being either the    particle horizon $L_\mathrm{p}$ 
or the future event horizon $L_\mathrm{f}$, we obtain 
 \begin{equation}
\label{H8d}
\frac{m}{a} =  \frac{d}{dt} \left( \frac{1}{a} \sqrt{ \frac{c^2}{H^2} 
 - \frac{1}{\Lambda_\mathrm{UV}^2}} \right) \, ,
\end{equation}
or equivalently  
\begin{equation}
\label{H8e}
\dot H = - \frac{H^3}{c^2} \left\{m \sqrt{ \frac{c^2}{H^2}  - 
\frac{1}{\Lambda_\mathrm{UV}^2}} 
+ H \left( \frac{c^2}{H^2} - \frac{1}{\Lambda_\mathrm{UV}^2} \right) \right\} \, .
\end{equation}
As we will soon see, the above simple modification caused by $\Lambda_\mathrm{UV}$ has a 
crucial effect in the inflation realization, namely it can cause a successful exit from 
the de-Sitter solution (\ref{dS1}) obtained for $\Lambda_\mathrm{UV}\rightarrow\infty$, 
and most importantly it can lead to inflationary predictions in perfect agreement with 
observations.

The general solution of (\ref{H8e}), for $\Lambda_\mathrm{UV}$ not being equal to 0 or 
infinity, can be written in an implicit form as
{\small{\begin{eqnarray}
&&
\!\!\!\!\!\!\!\!\!\!\!\!\!
\frac{\sqrt{c^2-\frac{H^2}{\Lambda_\mathrm{UV}^2}}}{
(c^2-1)^{\frac{3}{2}} \Lambda_\mathrm{UV}^2 \sqrt{H^2-c^2\Lambda_\mathrm{UV}^2}}
\tan^{-1}\left[
\frac{mH}{\sqrt{c^2-1} \sqrt{H^2-c^2\Lambda_\mathrm{UV}^2}  } \right]
\nonumber\\
&&
\!\!\!\!\!\!\!\!\!\!\!\!\!
+\frac{m\sqrt{c^2\!-\!\frac{H^2}{\Lambda_\mathrm{UV}^2}}\!-\!c^2}{
c^2(c^2\!-\!1)H\Lambda_\mathrm{UV} ^2}+
\frac{
\tanh^{-1}\!\left[
\frac{H}{\sqrt{c^2\!-\!1}\Lambda_\mathrm{UV}} \right]}
{ (c^2-1)^{\frac{3}{2}}\Lambda_\mathrm{UV}^3 }
 =\!-\!\frac{t}{c^2 \Lambda_\mathrm{UV}^2}\!+\!C_0,
 \label{generalsolution}
\end{eqnarray}}}
whose integration provides the scale factor evolution. Note that the above solution is 
real and well behaved for all $c>0$.
Hence, since the evolution of 
$H(t)$ is 
known, we can straightforwardly obtain the Hubble slow-roll parameters $\epsilon_n$ 
(with $n$ positive integer), defined as 
\cite{Kaiser:1994vs,Sasaki:1995aw,Martin:2013tda,Woodard:2014jba}
\begin{eqnarray}
\epsilon_{n+1}\equiv \frac{d\ln |\epsilon_n|}{dN},
\label{epsilonnn}
\end{eqnarray}
with $\epsilon_0\equiv H_{ini}/H$ and $N\equiv\ln(a/a_{ini})$  the e-folding number, 
and where $a_{ini}$ is the scale factor at the beginning of inflation and $H_{ini}$ the 
corresponding Hubble parameter  (inflation ends when $\epsilon_1=1$). Thus, we can 
calculate the values of   
the inflationary observables, namely the scalar spectral index of the curvature 
perturbations $n_\mathrm{s}$, its running $\alpha_\mathrm{s} \equiv d n_\mathrm{s}/d 
\ln k$ with $k$ the absolute value of the wave number $\Vec{k}$,
the tensor spectral 
index $n_\mathrm{T}$ and the tensor-to-scalar ratio $r$, 
as \cite{Martin:2013tda}
 \begin{eqnarray}
 r &\approx&16\epsilon_1 ,
 \label{eps111bb}\\
 n_\mathrm{s} &\approx& 1-2\epsilon_1-2\epsilon_2  ,
\label{eps2222bb} \\
\alpha_\mathrm{s} &\approx& -2 \epsilon_1\epsilon_2-\epsilon_2\epsilon_3  ,
\label{eps333bb}\\
 n_\mathrm{T} &\approx& -2\epsilon_1  ,
\label{eps444bb}
\end{eqnarray}
where  the 
first three $\epsilon_n$ are straightforwardly extracted from (\ref{epsilonnn})   to be 
\begin{eqnarray}
\label{epsVbb}
&&\!\!\!\!\!\!\!\!\!\!\!\!\!\!
\epsilon_1\equiv-\frac{\dot{H}}{H^2}, 
\\
&&\!\!\!\!\!\!\!\!\!\!\!\!\!\!
\epsilon_2 \equiv  \frac{\ddot{H}}{H\dot{H}}-\frac{2\dot{H}}{H^2},
\label{etaVbb}\\
&&\!\!\!\!\!\!\!\!\!\!\!\!\!\!
\epsilon_3 \equiv
\left(\ddot{H}H-2\dot{H}^2\right)^{-1}
\nonumber\\
&&\!\!\!\!
\cdot\!\left[\frac{H\dot{H}\dddot{H}-\ddot{H}(\dot{H}
^2+H\ddot{H}) } { H\dot { H } }-\frac{2\dot{H}}{H^2}(H\ddot{H}-2\dot{H}^2)
\right]\!.
\label{xiVbb}
\end{eqnarray}

{We mention here that in principle in order to extract the exact expressions for the 
inflationary observables we should perform a full perturbation analysis for this specific 
holographic model, based on the detailed perturbation analysis of holographic dark energy 
performed in \cite{Li:2008zq}. Nevertheless, for the purpose of the present work  the 
approximate relations (\ref{eps111bb})-(\ref{eps444bb}), that hold for every scenario  as 
long as $H(t)$ is known \cite{Martin:2013tda}, are adequate in order to provide the 
approximate values for the inflationary observables.

Relations (\ref{eps111bb})-(\ref{eps444bb}) are very useful, since they allow for a
comparison of the predictions of holographic inflation  with observations. 
In Fig.~\ref{rns} we present the estimated tensor-to-scalar ratio of the specific 
scenario for four parameter choices and for e-folding numbers varying between $N=50$ and 
$N=60$, on top of the 1$\sigma$ and 
2$\sigma$ contours of the Planck 2018 results \cite{Akrami:2018odb}. As we observe, the  
agreement with observations is very efficient, and in 
particular well inside the 1$\sigma$ region. 
 \begin{figure}[ht]
\centering
\includegraphics[scale=.50]{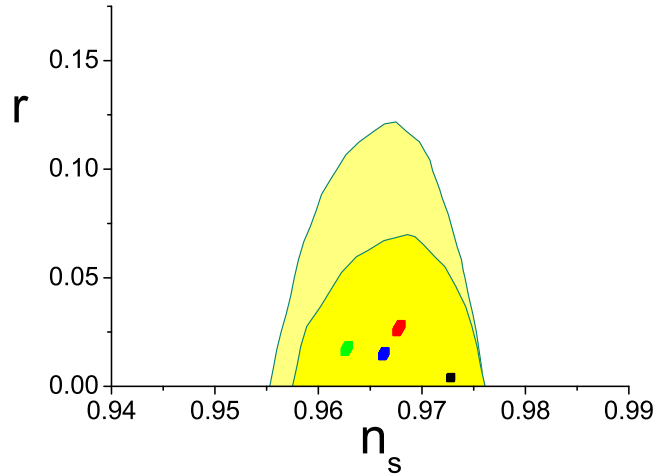}
\caption{{\it{ 1$\sigma$ (yellow) and 2$\sigma$ (light yellow) contours for Planck 2018
results (Planck $+TT+lowP$)  \cite{Akrami:2018odb}, on 
$n_{\mathrm{s}}-r$ plane.
Additionally, we present the predictions of holographic inflation for $m=-1$ (i.e. the 
future event horizon is used),  with 
$c=1.007$,$\Lambda_\mathrm{UV}=20$ (black points),
$c=1.009$,$\Lambda_\mathrm{UV}=19$ (red points),
$c=1.009$,$\Lambda_\mathrm{UV}=18$ (blue points), and
$c=1.01$,$\Lambda_\mathrm{UV}=18$ (green points), 
in units where $\kappa^2 = 1$, for e-folding number between $N=50$ and 
$N=60$ (for some of the cases the results for $N$ varying between $50$ and $60$ cannot be 
distinguished at the resolution scale of the figure).
}}}
\label{rns}
\end{figure}

In the scenario of holographic inflation examined above, the differential equation 
(\ref{H8e}) allows to eliminate the time-derivatives of $H$ in terms of $H$ in 
(\ref{epsVbb})-(\ref{xiVbb}), and therefore extract analytical and quite simple 
expressions for the inflationary observables.
In particular, doing so we find that
\begin{eqnarray}
\label{rsol}
&& r=16\left(
 1-\frac{H^2}{c^2 \Lambda_\mathrm{UV}^2}+\frac{m \sqrt{c^2-    
\frac{H^2}{\Lambda_\mathrm{UV}^2} }   }{c^2}
 \right),  \\
&&\label{nssol}
 n_\mathrm{s}=-1-\frac{2m}{\sqrt{c^2-\frac{H^2}{\Lambda_\mathrm{UV}^2}    
}}-\frac{2H^2}{c^2\Lambda^2_\mathrm{UV}},\\
&&
\label{assol}
\alpha_\mathrm{s} =
\frac{mH^4\left(m+ \sqrt{c^2-    
\frac{H^2}{\Lambda_\mathrm{UV}^2} }  
\right)}{c^4 \Lambda^2_\mathrm{UV} (c^2 \Lambda^2_\mathrm{UV}  -H^2)},
\\&& n_\mathrm{T} =-\frac{r}{8}.
\end{eqnarray}
Hence, we can now eliminate $\frac{H^2}{\Lambda_\mathrm{UV}^2}$ between 
(\ref{rsol}),(\ref{nssol}) and between (\ref{rsol}),(\ref{assol}), and obtain the 
relation of $ n_\mathrm{s}$ and $\alpha_\mathrm{s} $  in terms of $r$, namely
\begin{equation}
\label{rnsrelation}
 n_\mathrm{s}=-3+\frac{r}{8}+\frac{2-l\sqrt{c^2 r +4}}{2c^2}-\frac{8m}{\sqrt{c^2 r 
+8-4l\sqrt{c^2r+4}}},
\end{equation}
and
\begin{eqnarray}
\label{ralphasrelation}
&&
\!\!\!\!\!\!\!\!\!\!\!\!\!\!\!
\alpha_\mathrm{s} =\frac{m\left[
8+c^2(r-16)-4q\sqrt{c^2r+4}
 \right]^2  }
 {64 c^4  \left(
 8+c^2r-4q  \sqrt{c^2r+4}
 \right)}\nonumber\\
 &&
 \cdot
 \left(  
 4m+\sqrt{8+c^2r-4q  \sqrt{c^2r+4}   }
 \right) ,
\end{eqnarray}
where $l=\pm1$ corresponds to two solutions branches.
Interestingly enough, we observe that as long as  $\Lambda_\mathrm{UV}$ is finite 
and non-zero it does not appear in the above relations (\ref{rnsrelation}) and 
(\ref{ralphasrelation}), although its effect was crucial in generating a new solution 
branch that did not exist in the case where $\Lambda_\mathrm{UV}$ in (\ref{H8c}) was 
absent (namely when $\Lambda_\mathrm{UV}\rightarrow\infty$, i.e. when there is no 
Ultraviolet cutoff). Of course it does appear in the solution for $H$ (relation 
(\ref{generalsolution})), and hence along with $c$ it determines the duration of 
inflation and the 
e-folding number, and thus the bounds of the above parametric curves.

However, the most significant feature is that relation (\ref{rnsrelation}) leads to 
$r$ and  $n_\mathrm{s}$ values in perfect agreement with observations, as long as one 
chooses suitably the holographic parameter $c$. In particular, one can deduce that the 
most 
interesting case is when $m=-1$ (i.e. the future event horizon is used) and when $q=-1$. 
In this case, as one can see from Fig. \ref{rns2},
with $c$ slightly larger than 1 we can obtain a $n_\mathrm{s}$ inside its 
observational bounds and $r$ adequately small. Finally, 
calculations of  
$\alpha_\mathrm{s} $ using (\ref{ralphasrelation}) result  to typical values of the order 
of $10^{-7}$, and thus well inside the observational bounds  \cite{Akrami:2018odb}. These 
are the main result of the present work 
and reveal the capabilities of holographic inflation, since it is well-known that the 
majority of inflationary models cannot result to adequately small $r$. 
 \begin{figure}[ht]
\centering
\includegraphics[scale=.50]{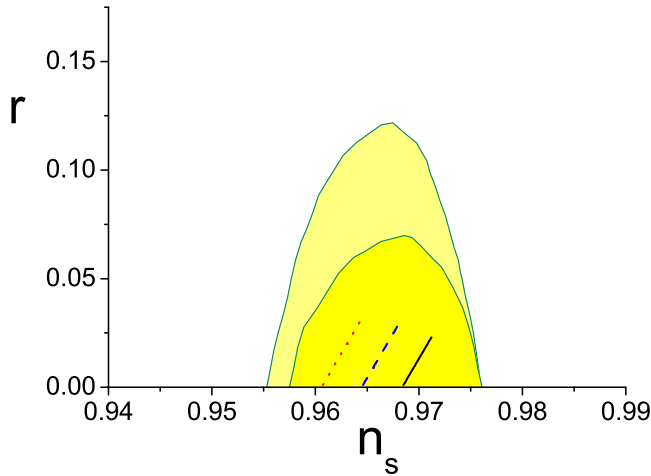}
\caption{{\it{  
1$\sigma$ (yellow) and 2$\sigma$ (light yellow) contours for Planck 2018
results (Planck $+TT+lowP$)  \cite{Akrami:2018odb}, on 
$n_{\mathrm{s}}-r$ plane.
Additionally, we present the analytical predictions of holographic inflation, namely the 
parametric curve (\ref{rnsrelation}), for  $m=-1$   and   $q=-1$, and   for $c=1.008$ 
(black-solid curve), for $c=1.009$ (blue-dashed curve), and for $c=1.010$ (red-dotted 
curve).   $\Lambda_\mathrm{UV}$ does not explicitly determine the form of the curves, 
however since it affects the duration of inflation and the e-folding number it 
implicitly determines their bounds (we have considered $\Lambda_\mathrm{UV}$ to vary 
between 1 and 100 in units where $\kappa^2 = 1$ and the e-folding number to vary between 
$N=50$ and 
$N=60$). 
}}}
\label{rns2}
\end{figure}

We close this analysis by making a comment on reheating, which is a necessary phase that 
should follow the inflationary one. In our model, the reheating can be obtained in a 
similar way to Starobinsky $R^2$ inflation, namely driven by scalaron 
(composite) degree of freedom. Hence, with the addition of this mechanism, we 
can obtain the transition to the post-inflationary thermal history of the universe.\\

\emph{Generalized scenarios} --  
In this section we apply the holographic principle at early times, however considering 
extended Infrared cutoffs. Such extensions have been applied in the late-time universe, 
resulting in generalized holographic dark energy  \cite{Nojiri:2017opc}. In particular, 
in these holographic constructions one considers a general Infrared cut-off 
$L_\mathrm{IR}$, which could be 
a function of both   $L_\mathrm{p}$ and $L_\mathrm{f}$  \cite{Elizalde:2005ju} and their 
derivatives, or 
additionally of the Hubble horizon and its derivatives as well as of the scale factor  
\cite{Nojiri:2017opc}, namely
\begin{equation}
\label{generalLIR}\!\!\!
L_\mathrm{IR}\!=\!L_\mathrm{IR}\!\left(L_\mathrm{p}, \dot L_\mathrm{p}, \ddot 
L_\mathrm{p}, \cdots, L_\mathrm{f}, \dot L_\mathrm{f}, 
\ddot 
L_\mathrm{f}, \cdots, a,
H, \dot H, \ddot H, \cdots\! \right)\! .
 \end{equation}
Hence, applying the above general Infrared cutoff at the early universe gives as 
enhanced freedom to obtain a successful inflationary realization.

Without loss of generality we consider the following specific example. We start by 
considering  an Infrared cutoff of the form
\begin{equation}
\label{H10}
L_\mathrm{IR} = - \frac{1}{6\alpha{\dot H}^2 a^6}  \int^t dt a^{6} \dot H \, ,
\end{equation}
with $\alpha$ the model parameter. Inserting this expression into 
the inflationary Friedmann equation (\ref{H2}) and taking $c=1$ for simplicity, leads to 
the differential equation 
\begin{equation}
\label{H11}
3 H^2   = \alpha \left( - 108 H^2 \dot H + 18 {\dot H}^2 - 36 H \ddot H \right) \, .
\end{equation}
Having in mind that the Ricci scalar in FRW geometry is just $R=6(2H^2+\dot{H})$, the 
above differential equation can be re-written in the form 
\begin{equation}
\label{H12}
 \frac{F(R)}{2}=  3\left(H^2 + \dot H\right) F'(R)
 - 18 \left( 4H^2 \dot H + H \ddot H\right) F''(R)  ,
\end{equation}
with $ F(R) = R + \alpha R^2 $.
Equation (\ref{H12}) is just   the first Friedmann equation of the $R^2$-gravity 
\cite{Starobinsky:1982ee,Capozziello:2002rd,Carroll:2003wy,Nojiri:2003ft}
(see \cite{Nojiri:2017ncd,Capozziello:2011et} for 
 reviews in $F(R)$ gravity) in the absence of matter.
 Therefore, the scenario of holographic inflation, under the generalized Infrared cutoff  
(\ref{H10}) can reproduce  Starobinsky $R^2$ inflation \cite{Starobinsky:1982ee}, which 
is known to lead to inflationary observables in a very good agreement with observations 
\cite{Akrami:2018odb}. 

In similar lines, one can consider other generalized Infrared cutoffs in order to obtain 
a correspondence with other geometrical inflationary models, such as Gauss-Bonnet and 
$f(G)$ inflation \cite{Kanti:2015pda}, $f(T)$ inflation, etc. These capabilities act as 
an additional advantage in favour of 
generalized holographic inflation (see also \cite{Binetruy:2014zya}).\\

\emph{Conclusions} -- 
In this work we applied the holographic principle at the early universe, obtaining an 
inflation realization of holographic origin. Although holographic energy density has been 
well studied at late times, giving rise to holographic dark energy, up to now it had not 
been incorporated at early times, although such consideration has equal footing, and 
moreover despite the fact that the decrease of the horizons at early times naturally 
increases holographic energy density at inflationary scales.
 
Taking as Infrared cutoff the particle or future event horizons, and adding a simple 
correction due to the Ultraviolet cutoff, whose role is non-negligible at the high energy 
scales of inflation, we resulted in a holographic inflation scenario that is very 
efficient in incorporating inflationary requirements and predictions. In particular, we 
first extracted analytically the solution of the Hubble function  in an implicit 
form, which can give a scale factor evolution of the desired e-foldings.

Furthermore, we 
analytically calculated the Hubble slow-roll parameters and then the inflation-related 
observables, such as the scalar spectral index and its running, 
the tensor-to-scalar ratio, and the tensor spectral index, which were found to follow 
simple expressions. Confronting the predictions with   Planck 2018 observations, we 
showed that the agreement is   perfect, and in particular 
deep inside the 1$\sigma$ region. Additionally, we found that with
$n_\mathrm{s}$ being inside its observational bounds $r$ can be adequately small.
  These are the main result of the present work and reveal the capabilities of holographic 
inflation, since it is well-known that the majority of inflationary models cannot lead 
to adequately small $r$. 
 
Finally, we constructed generalizations of holographic inflation which are based on 
extended Infrared cutoffs. Under these considerations we showed that we can reconstruct 
Starobinsky   inflation, which is also known to be in a very good agreement with the 
data, as well as   obtain a correspondence with other inflationary scenarios of 
geometrical origin. In summary, holographic inflation proves to have a very interesting 
phenomenology, and thus it is a good candidate for the description of the early universe.

\begin{acknowledgments} 
This work is partially supported  by MEXT KAKENHI Grant-in-Aid for 
Scientific Research on Innovative Areas gCosmic Accelerationh No. 15H05890 (S.N.) 
and the JSPS Grant-in-Aid for Scientific Research (C) No. 18K03615 (S.N.), 
and by MINECO (Spain), FIS2016-76363-P (S.D.O). 
\end{acknowledgments}

\end{document}